\documentstyle[aps,epsf,preprint]{revtex}
\tighten

\newcommand{\nnb}{\nonumber}
\newcommand{\bea}{\begin{eqnarray}}
\newcommand{\eea}{\end{eqnarray}}

\newcommand{\barl}{\begin{array}{rl}}
\newcommand{\barr}{\begin{array}{rr}}
\newcommand{\ball}{\begin{array}{ll}}
\newcommand{\ea}{\end{array}}
\begin{document}
\draft

\title{Multijet events and parton showers}

\author{
F.~Krauss$^{1,2}$\thanks{E-mail: krauss@theory.phy.tu-dresden.de},
R.~Kuhn$^{1,2}$\thanks{E-mail: kuhn@theory.phy.tu-dresden.de},
G.~Soff$^{1}$}

\address{$^1$ Institut f\"ur Theoretische Physik, TU Dresden, 
         01062 Dresden, Germany,\\
         $^2$ Max--Planck--Institut f{\"u}r Physik komplexer Systeme,
             01187  Dresden, Germany}
\maketitle

\begin{abstract}
We propose a general approach for the description of multijet events
in the framework of QCD event generators. We introduce a new algorithm 
to match parton showers and arbitrary matrix elements 
for the production of any number of jets via the strong or 
electroweak interaction. We developped a new parton cascade code capable
to deal with multijet events at LEP energies and beyond.
\end{abstract}
\begin{center}
{ PACS numbers 13.65.+i}
\end{center}
%
% INTRO
%
The formation of jets in $e^+e^-$ annihilations provides an interesting 
testing ground for many areas in modern particle physics. In the past 
years especially the production of four jets has drawn considerable 
attention for several reasons.

First, the interpretation of these events by means of the processes
\bea\label{ME1}
e^+e^-\to q\bar qq'\bar q'\;,\;\;\; e^+e^-\to q\bar qgg
\eea
respectively, allows for a test of the symmetry group underlying the strong 
interaction \cite{4JETth} via their colour factors $C_A$, $C_F$ and $T_F$
\cite{pQCD}. 
Second, since the corresponding cross sections depend sensitively on 
the number of flavours active at the relevant scale, effects of new 
physics might be constrained by their precice determination \cite{GLUINO}.
Third, four jet events represent a serious background to the
production and decay of pairs of electroweak gauge bosons
and of the Higgs boson of the Standard Model \cite{WWback}.
Last but not least, these processes might provide additional signatures for 
new particles. 

This phenomenological impact clearly signals the need for an adequate and 
thorough description of four and more jet events and a proper implementation 
into a multiple purpose Monte Carlo event generator. 

In principle, multijet topologies like the ones exemplified above, Eq.
(\ref{ME1}), are well described by the corresponding matrix elements evaluated 
in the standard way of summing and squaring amplitudes. Unfortunately the 
final states accessible by this method are not the hadrons 
observed experimentally
but partons. Therefore some hadronization scheme has to be used with usually
scale dependent parameters. Thus perturbative jet evolution as modelled
by the parton shower approach is an important feature of event generators
since it allows for the construction of a process independent infrared
scale of the order of a few $\Lambda_{\rm QCD}$ defining the onset of 
hadronization and consequently guaranteeing its universality. This explains 
the need for a general procedure to match all possible matrix elements 
for the various channels of jet production and the parton shower
\cite{MATCHJET,MATCHING,4MATCH}. 

%
% ME
%
Before discussing in some detail the matching procedure used within the new 
event generator APACIC++ we would like to summarize the relevant 
characteristics of matrix elements and parton showers. 

As indicated above the matrix elements are related to specific processes, in 
our case to processes like the one of Eq. (\ref{ME1}), describing the hard 
production of a number of partons. They can be evaluated perturbatively
in the standard way, results for the production of four jets via the strong 
interaction exist in Leading \cite{4JETLO} and Next--to Leading Order
\cite{4JETNLO}. Here we do not want to comment on the pecularities related to 
the mutual cancellation of virtual and real divergencies occuring in the 
perturbative treatment of the matrix elements. However, we would like to 
stress, that the real divergencies 
related to soft and collinear emissions can be avoided by the requirement
to describe exclusive processes, i.e. the production of a number of jets 
instead of partons. Popular schemes used for the definition of jets are
the JADE-- \cite{JADE} and DURHAM--algorithm \cite{DURHAM}.  
At this point it should be noted that the matrix elements describe on--shell
particles in the final state lacking the possibility to radiate further
and produce a jet. 

%
% PS
%
The jet evolution is commonly modelled by the parton shower approach.
Within the presentation of the matching algorithm employed by our code
APACIC++ we refer to a parton shower ordered by virtualities
with an angular veto on growing branching angles \cite{ANGORD} to 
account for the proper internal jet structure. Additionally, we take
all partons massless. However, an extension to a parton shower 
ordered by angles \cite{MLLA} and massive partons is straightforward.
Within this framework, the virtualities of the radiating partons play 
the role of an order parameter organizing the multiple emissions of 
the shower approach via the corresponding Sudakov form factor \cite{SUDAKOV}
\begin{equation}
\Delta(t)=\exp\left[-\int\limits_{t_0}^{t}
\frac{{\rm d}t'}{t'}\int{\rm d}z\frac{\alpha_s(\mu(t',z))}
{2\pi}\hat P(z)\right]\,.
\end{equation}
Note, that the splitting function $\hat P(z)$ entering the Sudakov form 
factor is derived for massless on--shell particles in the final--state. 
To account for local four--momentum conservation one commonly changes the 
four momenta of the outgoing particles to correct for effects related to the 
virtual masses they pick up in their subsequent decays, for details see 
\cite{PYTHIA}. In this respect, the virtualities represent a minor 
perturbation of the branching kinematics. 

The construction of a parton shower relies on an expansion around the soft
and collinear limits of the emission and the angular ordering accounts for the
coherence of the shower. Taken together, both features allow for a 
construction of jets by means of subsequent independent parton emissions.
Even the inclusion of azimuthal correlations steming from the helicity 
structure involved in two subsequent emissions does not alter this picture 
drastically \cite{AZIMUT}. Therefore it should be obvious that the parton 
shower approach is not capable of modelling the complicated topological 
structure of multijet events and one has to rely on the matrix elements and 
match the parton shower.

According to the considerations above the problem to match the matrix elements
to the parton shower can be reformulated as the question of how to supply
the final--state partons of the matrix elements with virtual masses
in a sensible way.

%
% MATCHING
%
We now turn to the discussion of our matching procedure once a kinematical
configuration of initial partons, i.e. jets, is given. Our approach 
basically requires the production of a jet to be a hard process
modelled by the corresponding matrix elements and the jet evolution to be
governed by the parton shower. Following the reasoning of \cite{MATCHING} 
we therefore separate the scales for jet production and the onset of jet 
evolution appropriately. In difference to their procedure to define the 
separation scale by a suitable choice of a fixed $Q^2$ as the initial virtual 
mass of every parton giving rise to a jet, however, we prefer to use 
the corresponding Sudakov form factor to determine this virtuality, i.e.
the invariant mass of the resulting jet. The choice of the Sudakov form factor
to govern the virtual masses of the initial partons results in varying 
invariant masses steming from the hard process and a smoother transition of 
topologies involving different numbers of jets. Of course the number of jets
in a specific event can be chosen according to the perturbative rate.

Since the Sudakov form factor enters the parton shower in the form
\bea\label{Sud}
{\cal P}_{\rm no} = \frac{\Delta(t_{ij})}{\Delta(t_i)}
\eea
determining the probability of no resolvable parton emission between 
$t_{ij}$ and $t_i$ we have to construct the virtual masses 
$t_{ij}=(p_i+p_j)^2$ of the intermediate states $ij$ the outgoing partons 
$i$ and $j$ originate from. Equipped with this intermediate virtual masses we 
are able to construct the virtualities $t_i$ and $t_j$ of the initial partons 
$i$ and $j$ via the corresponding Sudakov form factor of Eq. (\ref{Sud}).
The task left is now to determine these intermediate virtualities.
In most cases, a number of Feynman diagrams contributes to the matrix element 
under consideration. Thus one has to define relative probabilities of the
various parton histories leading to different $t_{ij}$ allowing to chose
one of the histories by random.

The algorithm described above can be best illustrated by considering the
production of four jets as indicated in Eq. (\ref{ME1}). Seven Feynman 
diagrams contribute and in a first step one choses the flavours of the four 
partons by random according to the corresponding cross sections. This is in 
full analogy to the determination of the number of jets.

\begin{center}
\begin{figure}
\epsfxsize=6cm\epsffile{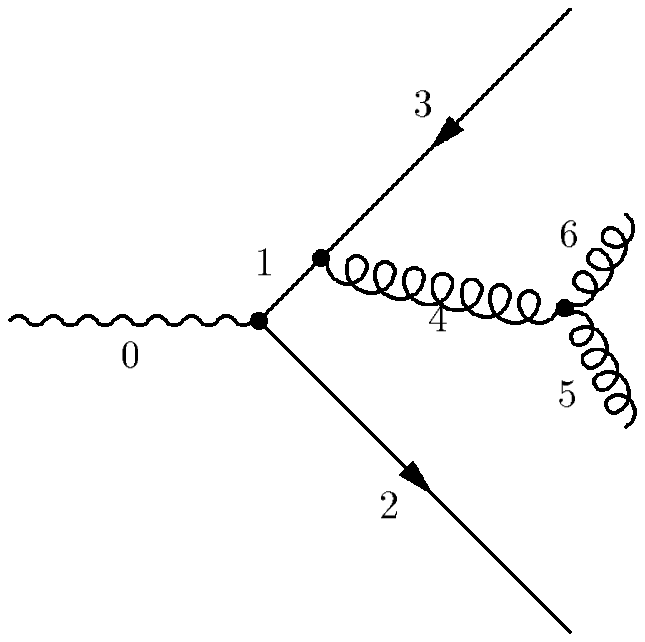}
\caption{\label{feyn}\footnotesize{Typical graph for 
$e^+e^-\to$ {\em four jets} at leading order.}}
\end{figure}
\end{center}

Let us assume in the following, that we have chosen the 
$q\bar qgg$--configuration and that we have constructed the momenta of the
outgoing partons according to the corresponding differential distribution. 
So we are now able to construct relative probabilities
of the five histories or diagrams $i$ using the appropriate amplitudes 
${\cal M}_i$ via
\bea
{\cal P}_i = \left|{\cal M}_i\right|^2.
\eea
Alternatively one might reconstruct
the individual parton histories in a fashion motivated by the parton shower
\cite{MATCHJET}. For example, in this approach the probability of the history
depicted in Fig. \ref{feyn} reads
\bea
{\cal P}={\cal P}_{1\to 34}{\cal P}_{4\to 56}=
\frac{1}{t_1}P_{qg}(z_{34})\,\,\frac{1}{t_4}P_{gg}(z_{56})\,,
\eea
where the $P(z)$ are the well--known splitting functions related to 
the branching processes involved, the $t_i$ are the squares
of the four momenta as given by 
\bea
t_1 &=&p_1^2 = (p_3+p_5+p_6)^2\,,\nnb\\
t_4 &=& p_4^2 = (p_5+p_6)^2\,,
\eea
and the $z_i$ are defined by
\bea
z_{bc} &=& z_{a\to bc}=\frac{t_a}{\lambda}\frac{E_b}{E_c} -
          \frac{t_a-\lambda+t_b+t_c}{2\lambda}\,,\nnb\\
\lambda &=& \sqrt{(t_a-t_b-t_c)^2-4t_bt_c}\,.
\eea
Having chosen one of the histories according to the relative probabilities
it is 
straightforward to construct the virtual masses of the intermediate states
previously denoted as $t_{ij}$. In our example, Fig. \ref{feyn}, this
amounts to calculate $t_4=p_4^2(p_5+p_6)^2$ and $t_1=p_1^2=(p_3+p_4)^2$.

Now the Sudakov form factor determines the virtual mass of the outgoing 
particle under the requirement that its decay does not produce any additional
jet. The kinematical distribution of the final state particles of the matrix 
element is corrected in the manner employed within the parton shower to 
conserve the total four momentum at this place, too. 

Two further remarks are in order here. First, it should be noted that our
matching procedure of parton showers and matrix elements relies on an internal
jet--clustering algorithm and yields sensible results for observables
related to jets as long as we produce them only perturbatively. Other
observables do not depend sensitively on this internal clustering scheme.
Second, to account for the appropriate jet--rates it should be mentioned
that so far the scale of the couplings and especially the scale of $\alpha_s$ 
within 
the matrix element expressions has not been fixed. We propose to use an
effective scale $s_{\rm eff}=s\kappa_s$ for the various coupling constants.
With a suitable choice of the scalefactor $\kappa_s$ rates for different 
numbers of jets at various energies can be accounted for simultaneously. 
This enables us to treat the QCD--production of varying numbers of jets on 
equal footing.

%
% Results
%
We have performed a comparison of a variety of observables describing
the event shapes of three-- and four jet--events at energies between
$91$ and $161$ GeV at the level of matrix elements, parton showers
and hadrons using PYTHIA \cite{PYTHIA}, HERWIG \cite{HERWIG} and our
event generator APACIC++. No initial state radiation was 
taken into account. We have used the matrix element expressions for the
production of up to four jets as provided by the code of \cite{DEBRECEN} and 
the hadronization scheme of \cite{STRING} as provided by JETSET linked
to our parton shower.

\begin{figure}
\begin{tabular}{cc}
\epsfxsize=8cm\epsffile{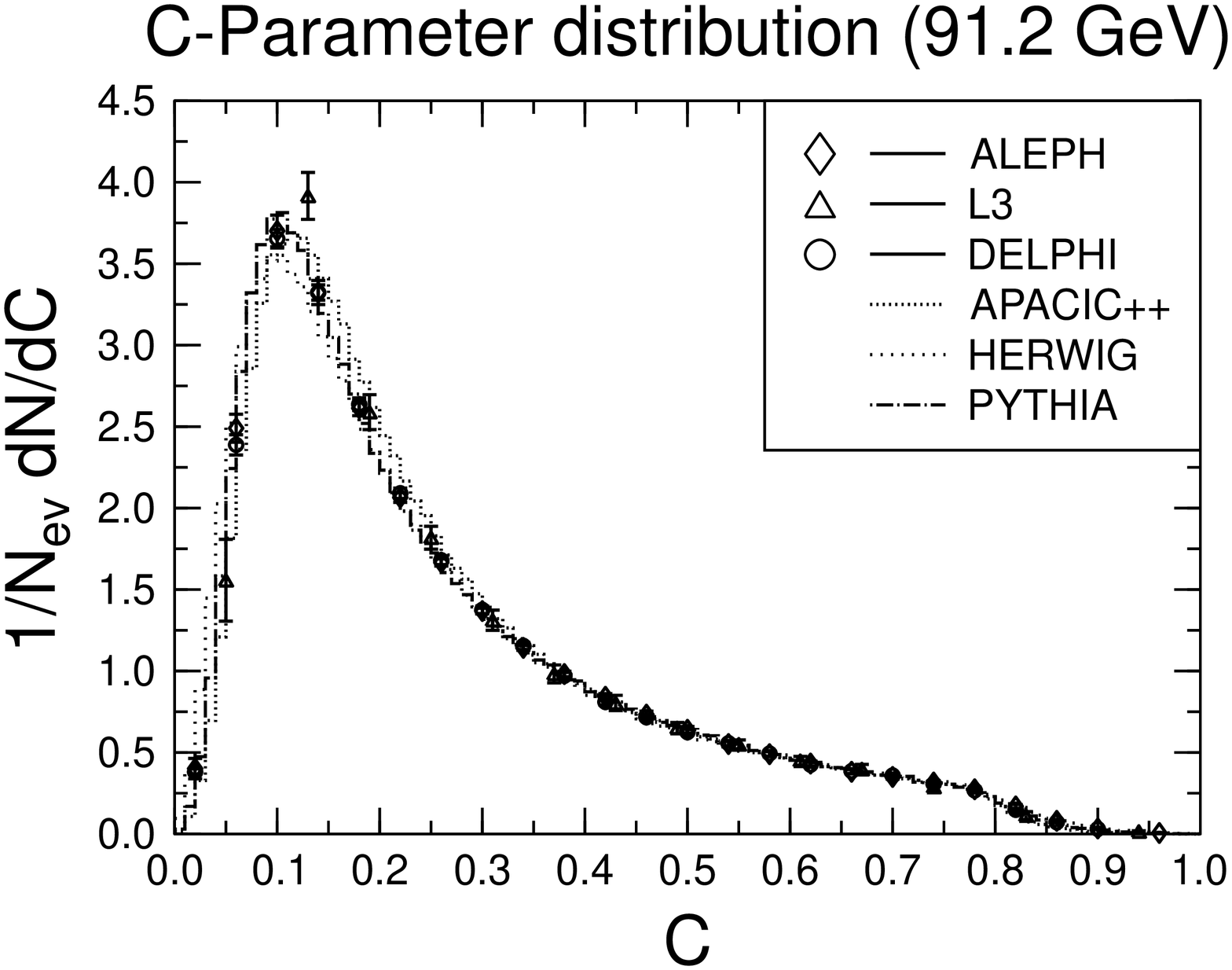}
\epsfxsize=8cm\epsffile{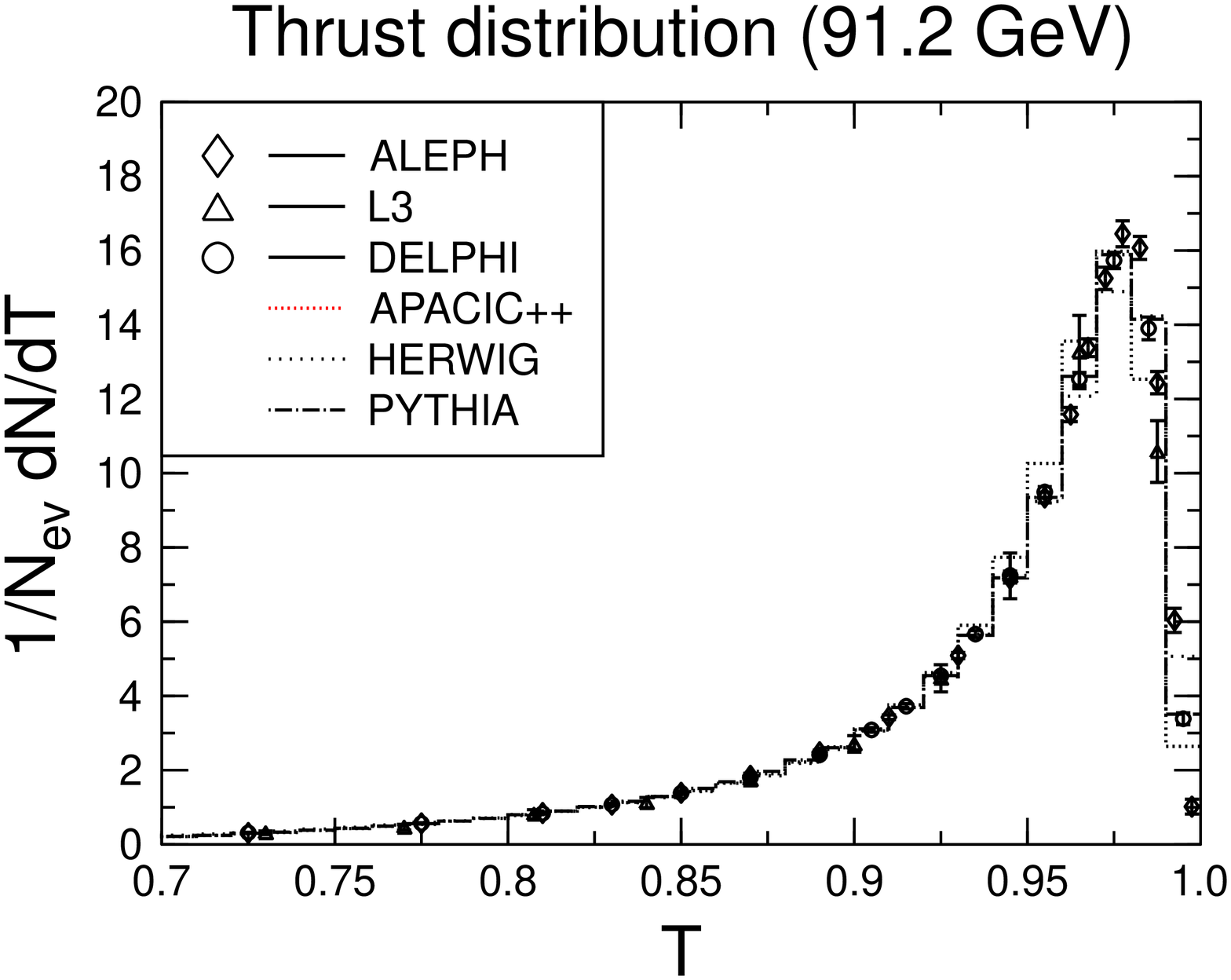}\\
\epsfxsize=8cm\epsffile{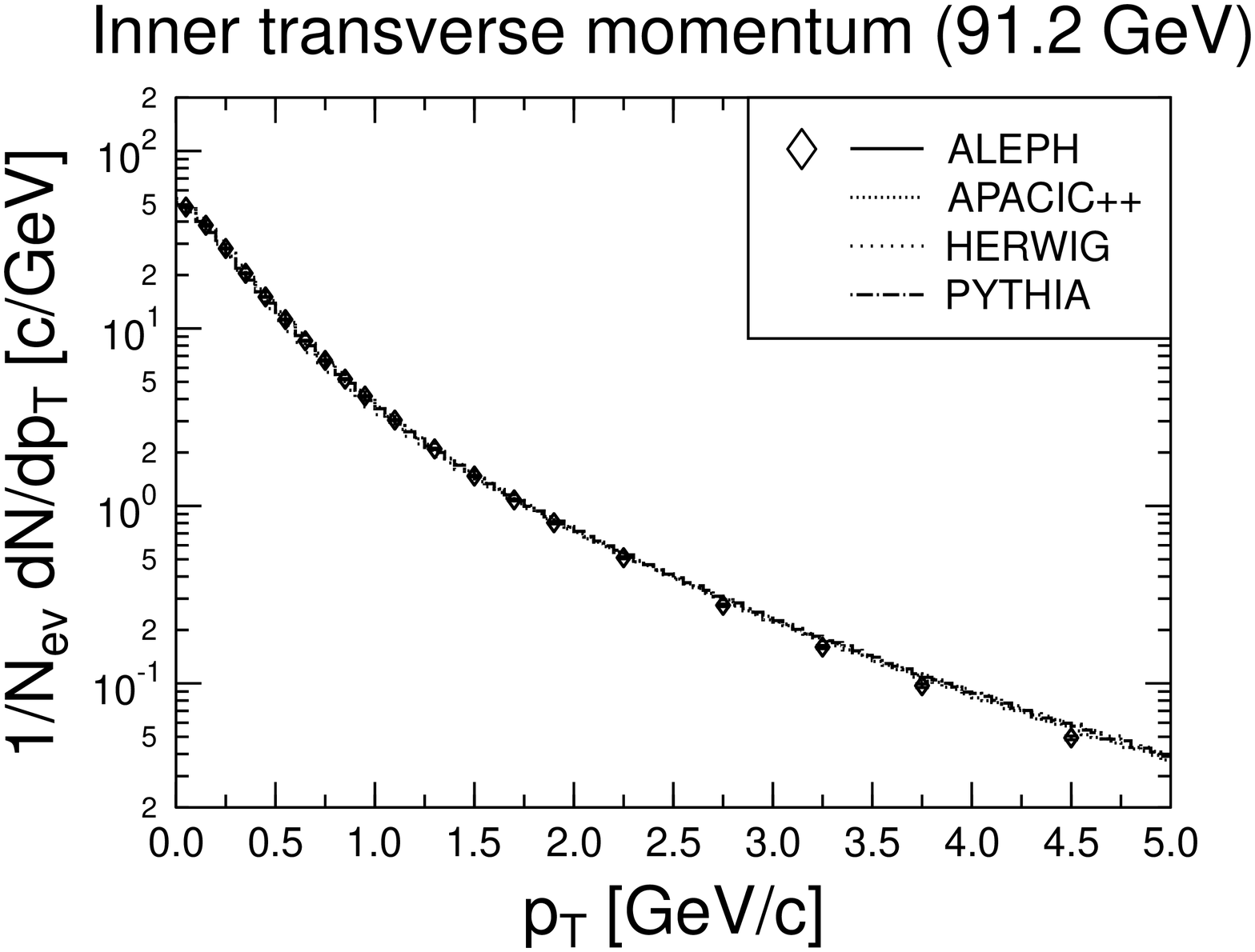}
\epsfxsize=8cm\epsffile{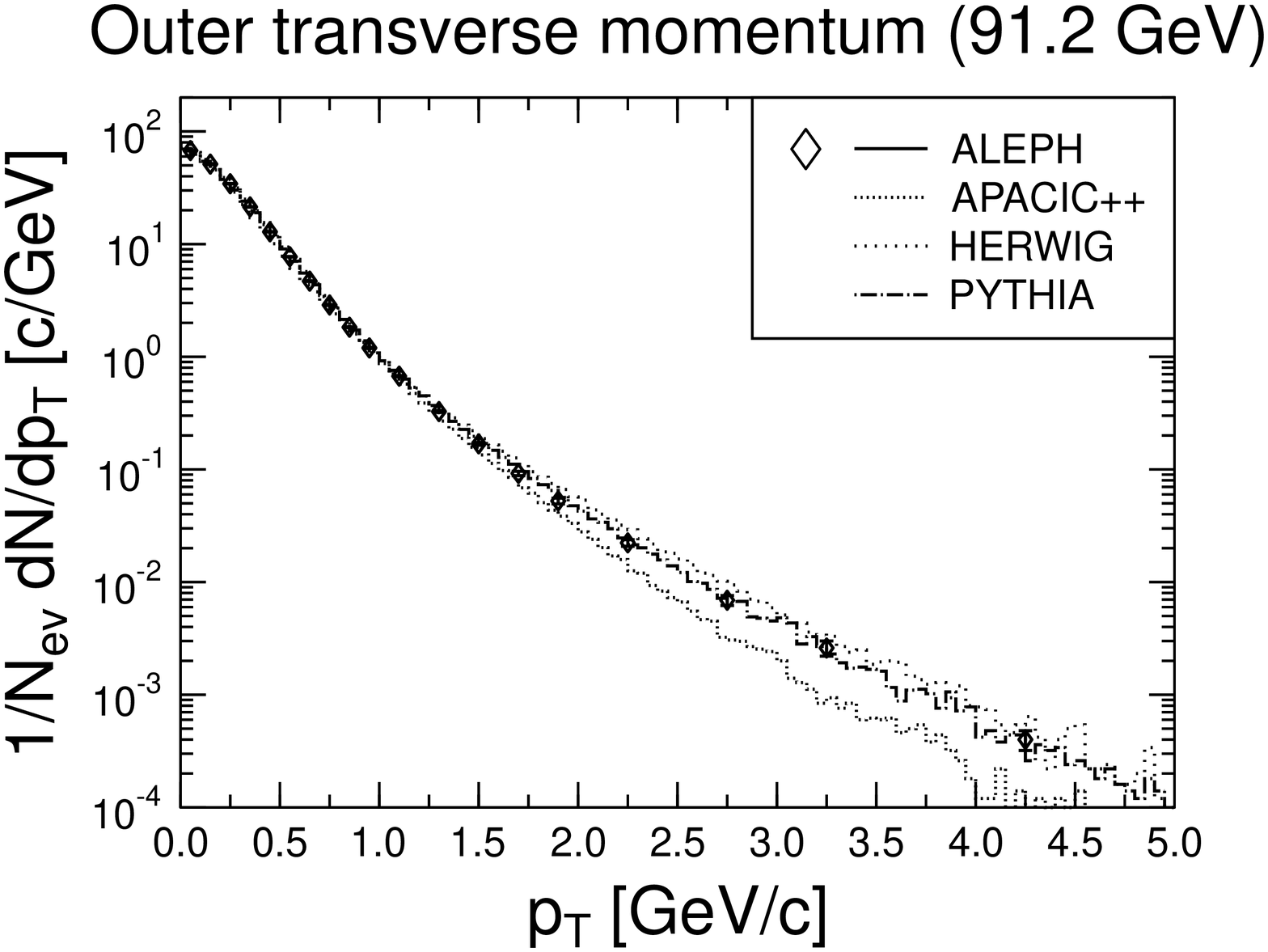}
\end{tabular}
\caption{\label{observs1}\footnotesize{Comparison of experimental data and 
event generators for a variety of event shape observables at the hadron level
at the $Z$--pole. 
For the hadronization the default parameters of
PYTHIA [15] and HERWIG [17] were used.
APACIC++ employed the Lund--String hadronization of PYTHIA.}}
\end{figure}

The results for most of the observables obtained with the different event 
generators coincide nicely with each other and with experimental data 
\cite{ALEPH91,DELPHI92,L392}. For a representative extract of various event
shape observables see Fig.\,\ref{observs1}. At this place it should be noted 
that we used the tuned parameters for the other event generators as given in 
\cite{Ham} representing the best choice to account for the experimental data.
In contrast the parameters of our code have not been fitted so far and we
used the unaltered hadronization parameters of PYTHIA. This somewhat lowers
the quality of the results produced by APACIC++. Nevertheless, the results of
APACIC++ are in fair agreement with data indicating that our approach to
match matrix elements and parton showers 
is perfectly capable to describe the interplay of various numbers
of jets as well as the overall features of $e^+e^-$ events.

However, the validity of our ansatz can be verified when
considering the topological structures of multijet events as exemplified
by four--jet events. Ordering the jets by their energies, 
$E_1\ge E_2\ge E_3\ge E_4$, typical observables describing these processes 
are the modified Nachtmann--Reiter--, the K{\"o}rner--Schierholz--Willrodt--
and the Bengtson--Zerwas--angle as well as $\alpha_{34}$ \cite{4JETth,4JETth2},
\bea\label{angledef}
 \theta_{\rm NR}^* &=& \angle(\vec{p_1}-\vec{p_2},\vec{p_3}-\vec{p_4})
 \,,\nnb\\
 \chi_{\rm BZ} &=& \angle(\vec{p_1}\times\vec{p_2},
                       \vec{p_3}\times\vec{p_4})\,,\nnb\\
\theta_{34} &=& \angle(\vec{p_3},\vec{p_4})
\,,\nnb\\ 
\Phi_{\rm KSW}^* &=& \angle(\vec{p_1}\times\vec{p_3},
                       \vec{p_2}\times\vec{p_4})\,,
\eea
where the last defintion holds for $|\vec{p_1}+\vec{p_3}|\ge
|\vec{p_1}+\vec{p_4}|$ and in the opposite case we interchange 
$\vec p_3$ and $\vec p_4$.

\begin{figure}
\begin{tabular}{cc}
\epsfxsize=8cm\epsffile{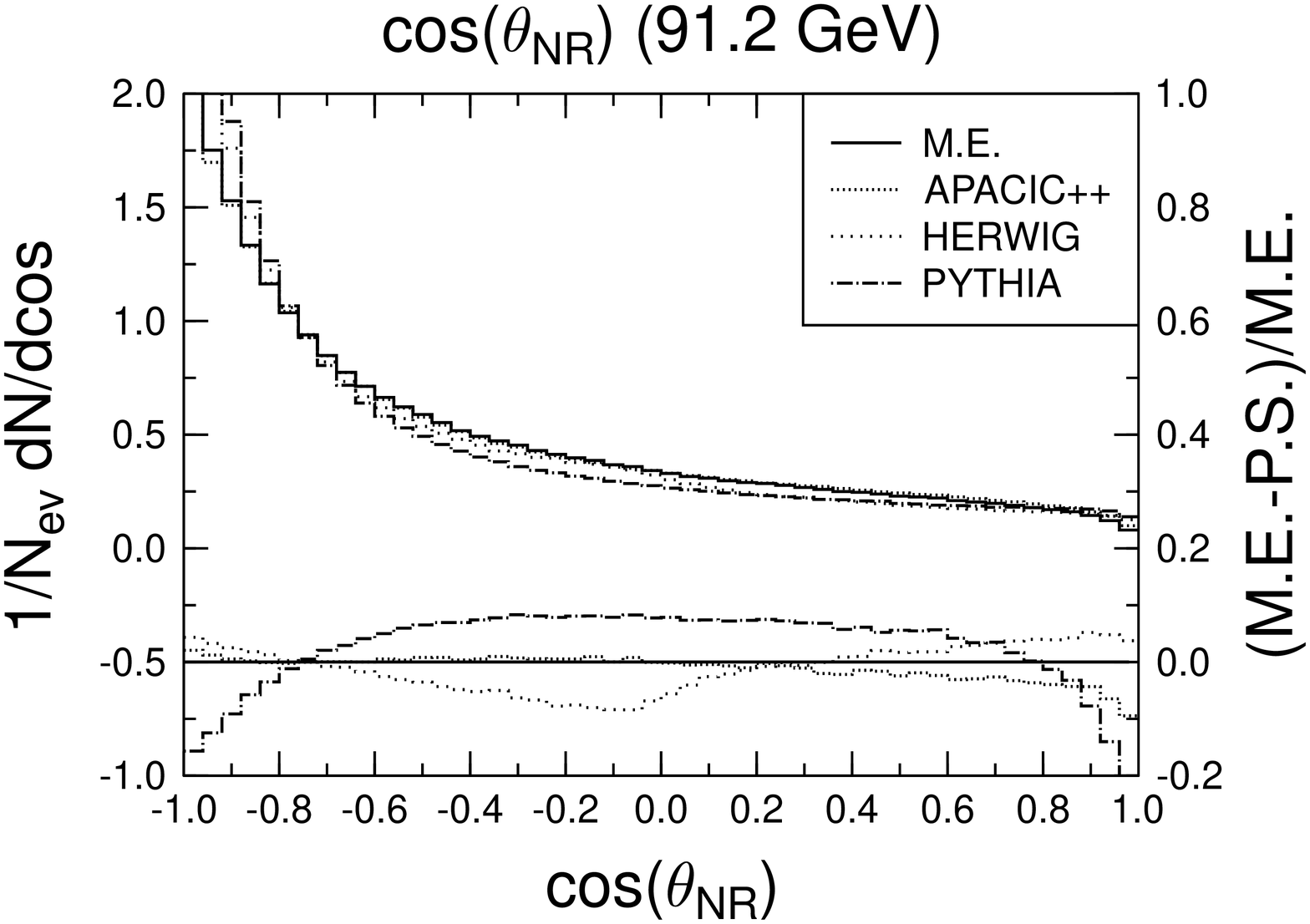}
\epsfxsize=8cm\epsffile{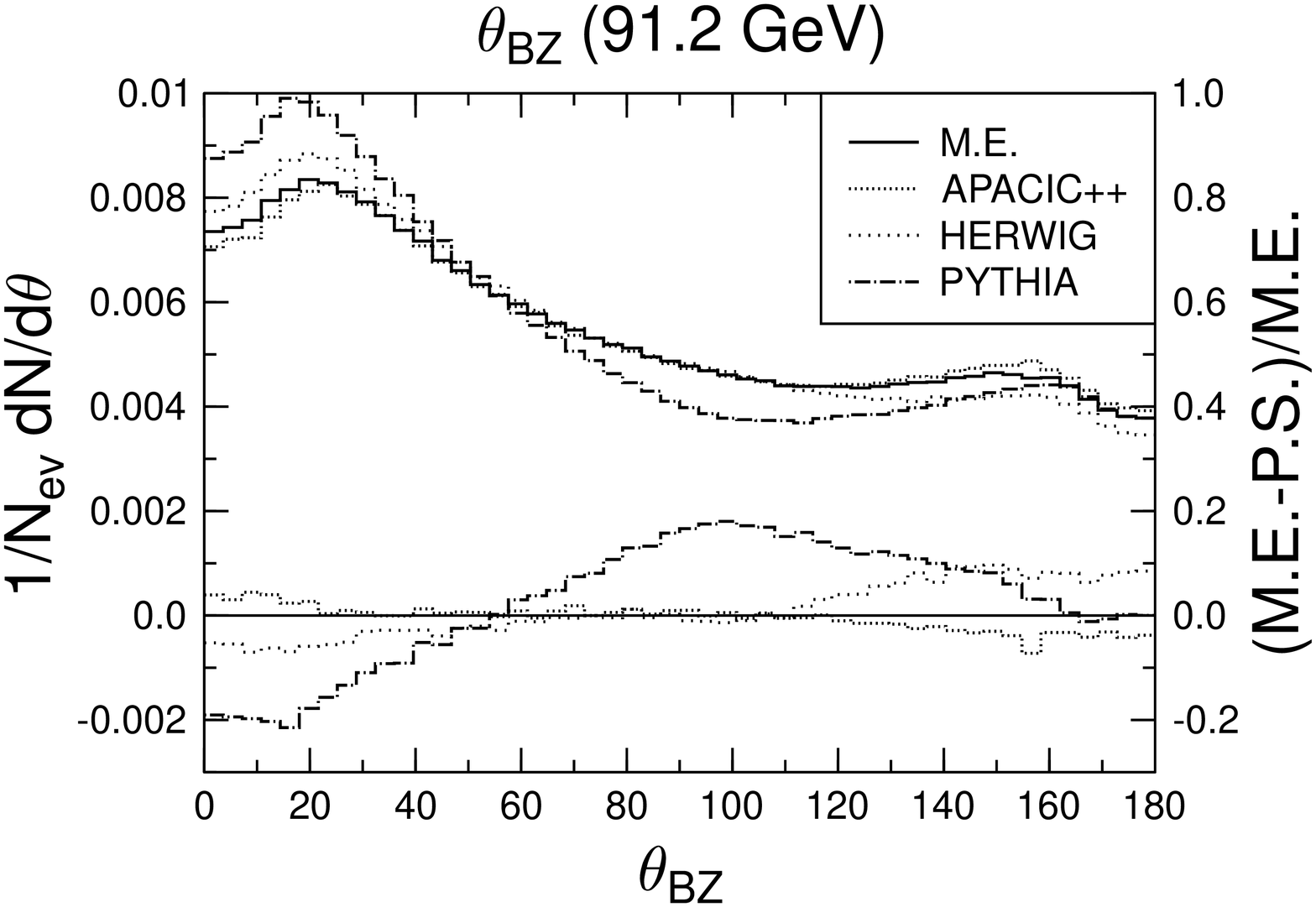}\\
\epsfxsize=8cm\epsffile{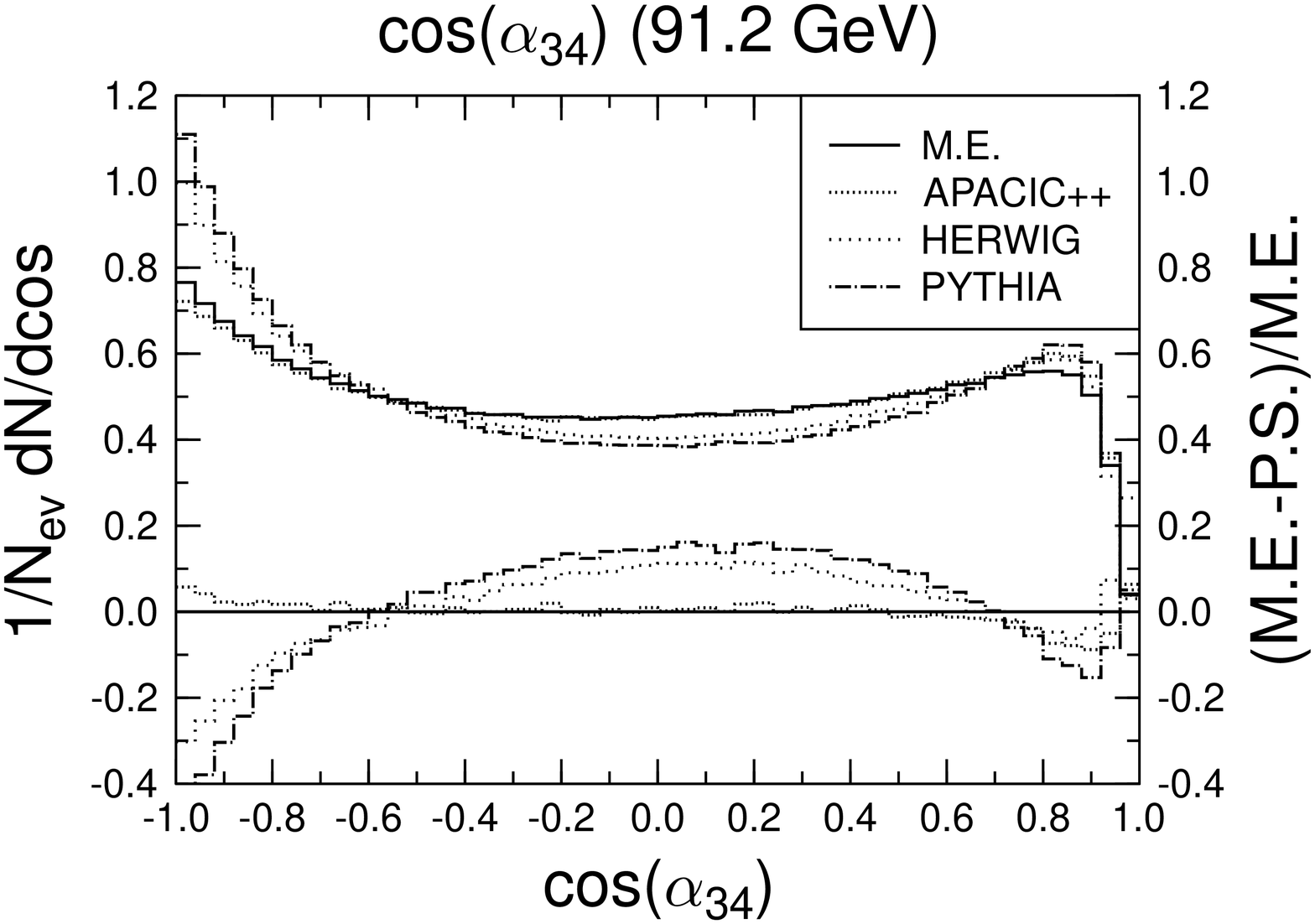}
\epsfxsize=8cm\epsffile{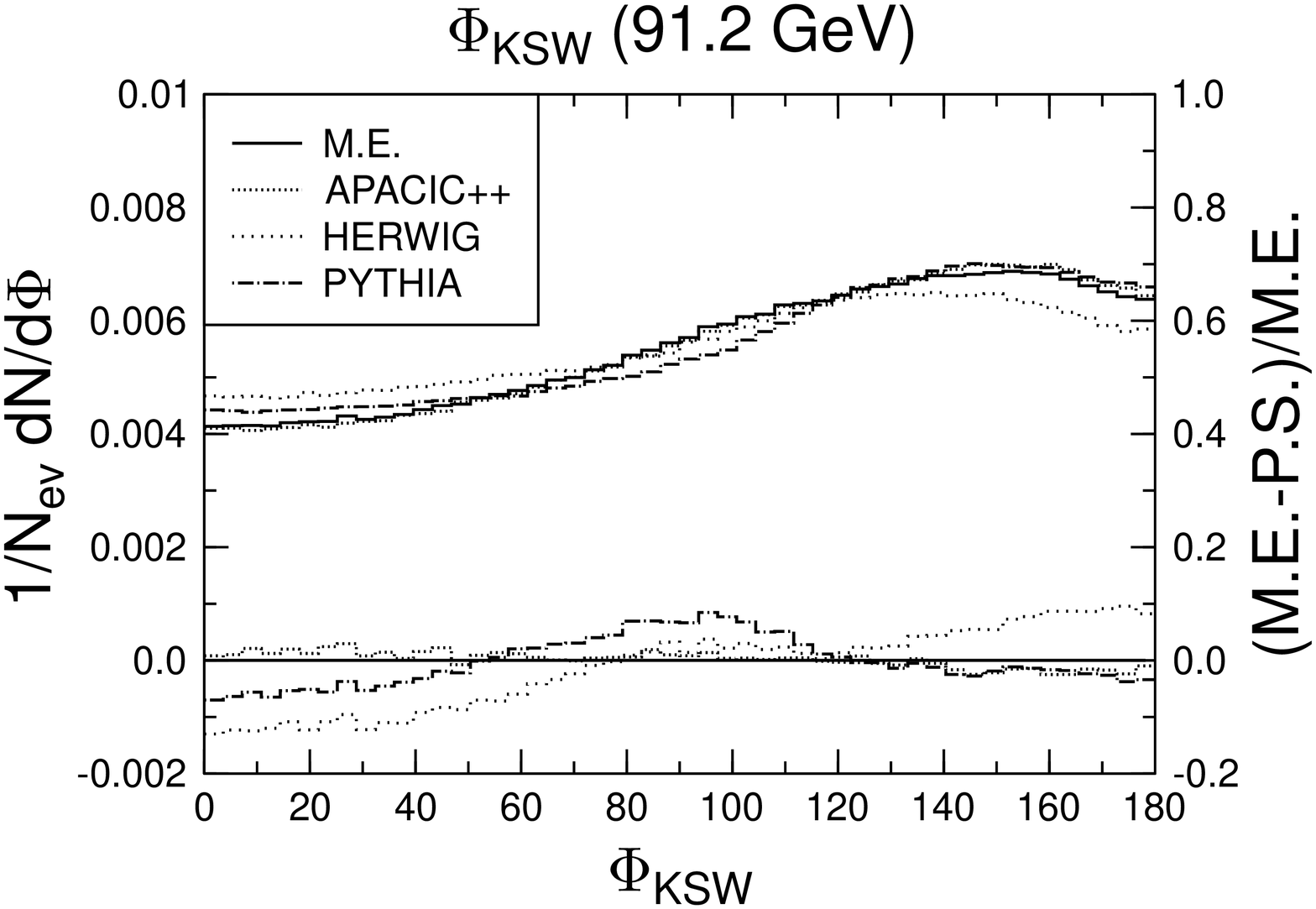}
\end{tabular}
\caption{\footnotesize{\label{angles}
         Distributions for the various angles given in 
         Eq. (\ref{angledef}) as described by the matrix elements supplied 
         by DEBRECEN [18] and the different event generators 
         at the $Z$--pole. 
         For the definition of jets the Durham--scheme with 
         $y_{\rm cut}=0.002$ was employed for all final states
         as well as for the matching of the matrix elements and
         the parton shower. The upper lines show the corresponding
         differential rates with respect to the numbers on the left 
         axis whereas the errors relative to the matrix element
         expression are are given by the appropriate lower lines with respect
         to the numbers on the right axis.}} 
\end{figure}

In Fig.\,\ref{angles} we show the angular distributions of the partons after 
the shower generated by the various event generators in comparison to the
distributions as given by the matrix element. Obviously APACIC++ is perfectly
capable to describe the four jet topologies in detail since the parton 
shower and the matrix element are matched appropriately. The only sizeable 
deviations 
of the jet distributions after the parton shower from the result as given by 
the matrix element alone are centered in the region of nearly collinear jets.
This is not too surprising, however, since the jet evolution widens
the initial partons to jets and therefore overshadows the jet definition
of two jets very close to each other.

In contrast, the two other event generators do not include an accurate 
matching procedure for events with more than three jets. Therefore, their 
parton shower results fail to describe any four-- or more jet specific 
topological structures.
So the large deviations for nearly the whole angular region under 
consideration merely reflect the fact that the parton shower alone is not 
capable to model the rich topological structure of multijet events.

We have proposed a general approach to match parton showers and arbitrary
matrix elements in the framework of QCD event generators. The results 
obtained with the help of our algorithm show a fair agreement with 
experimental data and matrix element expressions available and are almost
insensitive to internal parameters within reasonable regions. 
An extension to jets produced via the electroweak interaction is 
straightforward.

In this respect, our approach offers new possibilities to describe
some precision data concerning multijet events at LEP II and beyond.

\section*{Acknowledgements}

We would like to thank B. Ivanyi for lively and helpful discussions.
We gratefully acknowledge financial support by DFG, BMBF and GSI.

\begin{appendix}
\newpage
\end{appendix}

\end{document}